# Building the next pyramid

Dr. J. West*, G. Gallagher*, K. Waters**, Stephen Ward***, Tia Ward***
*Department of Chemistry and Physics, Indiana State University, Terre Haute, IN 47809
**Department of Physics, Michigan Technological University, Houghton MI 49931, graduate of Indiana State University
***Earth Pyramid, Global peace and Environmental Education Project, Leeds, West Yorkshire, UK

Abstract

The results of experimental tests of a novel method for moving large (pyramid construction size) stone blocks by rolling them are presented. The method is implemented by tying 12 identical rods of appropriately chosen radius to the faces of the block forming a rough dodecagon prism. Experiments using a 1,000 kg block show that it can be moved across level open ground with a dynamic coefficient of friction of less than 0.06. This value is a factor of five lower than that obtained for dragging the block, and the best values reported for dragging by others, at 0.3. The results are more dramatic than those obtained on smaller scale experiments on a 29.6 kg block, also reported here. For full scale pyramid blocks, the wooden "rods" would need to be posts of order 30 cm in diameter, similar in size to those used as masts on ships in the Nile.

# I. INTRODUCTION

Some of the pyramids in Egypt, architectural wonders, were built more than 4000 years ago. As a basis for the scale of the effort involved, it is noted that there are about 2.3 million blocks that make up the pyramid of Khufu.[1,2] Those blocks were quarried locally from limestone with average dimensions 1.27m x 1.27m x 0.71m and imposing mass of 2500kg.[1] The blocks, once quarried, were moved approximately 2 km to the construction site, and then into position, some to the top of the pyramid at an altitude of 138 m.[1-3] The method of construction of these impressive monuments has been a subject of intensive investigation, and speculation by scholars from a wide range of disciplines, and armchair engineers alike. In this paper we present the results of testing two methods of moving a 1,000 kg block. The topic has particular current relevance due to the project currently underway by an international coalition, Earth Pyramid,[4] to build a pyramid. The pyramid is to have 70 blocks along each base edge (for a total of 116,795 required blocks), and is to be built utilizing human-powered construction methods. The experiments reported in this paper were conducted for the purpose of assessing the usefulness of two of the methods for moving the large blocks which might be used in that project.

There have been a number of documentaries produced on the pyramids in general, and some have included large scale tests of possible construction methods. Still, the most often proposed method is that of placing the stones on a sled, and then dragging the sled via ropes across a prepared, possibly water lubricated, roadway. Of particular interest is a documentary episode produced by NOVA in which Mark Lehner and Roger Hopkins actually lead a team that built a small pyramid of approximately 10 m in height.[5] They tested moving the blocks by dragging them on sleds, or on sleds with some lubrication, over prepared roadways. The lubrication clearly had a large effect in that project, but tension gauges were not employed and so there is a lack of quantitative results from that test. In addition, not all of the pyramids were built on the banks of a major body of water, so that the transport of the lubricant in the form of the water might become a project as difficult as the moving of the blocks. As was used in this documentary, dragging the blocks on sleds is the default suggestion, and is seen in multiple illustrations from the ancient world, but in experiments testing the coefficient of friction in wet sand, Fall et. al.[6] found that the percent water content of the sand (about 5% was optimal) could decrease the "dynamic friction" $\mu_d$, by as much as 40%, but even under controlled conditions the lowest value reported was $\mu_d = 0.30$. This value sets the benchmark for the amount of work the crews would need to perform in moving the blocks to the building site. A very fit human using the major muscle groups can perform long term physical work at a rate of approximately 75 W.[7] In order to maintain a constant block speed of 0.5 m/s a block of mass 1,000 kg would require a work crew of roughly 20 individuals.

As an alternative to dragging large blocks, one can consider rolling the blocks. While it is clearly not efficient to roll the 4 sided blocks as they are, adding wooden rods to the surface can effectively increase the number of sides without a great increase in mass. The crew would pull on a rope wrapped around and passing over the top of the block. In this configuration, static friction acts **in the direction of the desired motion**, rather than opposing the motion. In effect the block and rope combination becomes a 2:1 pulley, though the pulley was not yet formally "known" to the Egyptians at that time.[1,3] In the experiments conducted on the 1,000 kg block, the results obtained using tension gauges are that dragging the stone on a prepared surface of water-lubricated sand gives $\mu_d = 0.25$, while rolling the block with the rods tied to the surface, and on a prepared surface gives $\mu_d = 0.054$, an improvement of roughly a factor of five! Note, due to the 2:1 pulley effect, the force required to maintain the motion of the block is only $(mg\mu_d)/2$, or about 2.5% of the weight of the block. The rods can be reused many times, and there is no need to transport large quantities of water for lubrication. The diameter of the rods required are consistent with the diameter of the masts used on ships in Egypt at that time,[8,9] the same ships that would be employed in moving granite blocks along the Nile.

Independent work appearing in the book *Engineering the Pyramids* by D. Parry[10] describes full scale experiments testing a related method for rolling large stones. Their tests used a 2,500 kg stone with dimensions 0.8 m x 0.8 m x 1.6 m. Wooden quarter circle "cradle runners" or "rockers" were attached, one to each end of each face of the block by simply wrapping rope around them, turning the square prism block into a cylinder. The rockers had been mentioned in other texts, but not in this context.[3] The stone was moved 80 meters on level ground (using 160 m of rope), and up a 1 in 4 slope, steeper than the 1:12 or 1:8 grade ramps that are in evidence at some of the pyramid sites.[1,3] The "block could be rolled at a fast walking pace by two or three men pushing it from behind." The block was moved much faster than the 4 meter / minute rate assumed by Smith, who had attempted to estimate the construction time and effort for the original pyramids, and a much smaller crew with much less effort than reported by Lehner and Hopkins.[4] The same block could be moved short distances up the 1:4 slope by 10 men, and was moved up the full 15 m ramp by 14, although a crew of 20 was more routinely used. Tension gauges were not used, but the number of men required to move the block suggests as a very rough estimate a value of $\mu_d = 0.2$. Perry refers to earlier tests by "an American engineer" named John Bush and carried out in Boston in 1980, using similar quarter circles are mentioned. Online sources place this work by Bush in 1977,[11] or 1978,[12] but no specifics about the experiments are supplied, and the results are not "published" as far as the authors can tell.

Another interesting variation on the idea of rolling the blocks, but again lacking actual values of forces involved, is to use catenary shaped wedges of wood to change the shape of the road, and leave the blocks free of attachment.[13] The results obtained are impressive, again at the level of teams of 2 or 3 people able to move the blocks. Modifying a roadway in such a manner is an idea that has been known for some time. A particularly nice example using a square wheeled tricycle is found on an SPS chapter website.[14] The method clearly works for blocks as well,[13] but requires the fabrication of many of the specially shaped road segments, and it is not clear that the road segments would be stable on inclines or in rainy conditions.

The remainder of the paper is organized as follows: Section II presents a short review of the effect by which friction is used to help move the blocks when rolling them, Section III describes the experimental methods, specifying the radius needed for the rods, the method of attaching the rods, and the experimental results. The conclusions are presented in Section IV.

**II. THEORY**

In dragging a block, it is clear that the work performed by the crew is lost as heating due to sliding friction. In rolling the block, most of the work performed in raising the center of mass from the face to a vertex is lost as it "falls" from the vertex onto the next face in an inelastic collision with the ground. However, not **all** of the mechanical energy is lost, and the fraction lost decreases as the number of faces on the polygon increases up to a limit where actual "rolling friction" would begin to dominate. The face collisions act, effectively, as a form of rolling friction, termed here the effective dynamic coefficient of friction $\mu_d$ for the polygon. If the crew moves at roughly a constant speed, they do work at a constant rate given by the "tension of rolling" (TR) times their speed counteracting those energy losses. (notice the speed of the crew is twice that of the block).

A complicating issue is that a "vertex tension" (TV) is required to get the polygon off of the resting face and up onto the first vertex to initiate rolling, and TV > TR. This is analogous to the larger value of the coefficient of static friction when compared to kinetic friction. An elementary analysis based on forces and torques shows that the ratio TV/TR decreases for a polygon of N sides with increasing N as $\tan(\pi/N)$. A large ratio is a bad thing for the crew, as it might mean additional members must be added, for the sole purpose of starting the rolling motion.

Rolling a cylinder with a rope wrapped over the top of the object is a well-known example/homework problem commonly addressed in first semester freshmen level physics courses.[15,16] It is used because of the "surprise" result that the force of friction is in the **same direction** as the applied tension, so that the acceleration a, of the cylinder is given by

$$a = \frac{4}{3}\frac{T}{m} < 4\mu_s g \qquad (1)$$

where T is the tension in the rope and m is the mass of the cylinder. The magnitude of the acceleration is **greater than** T/m, while limited by possible slipping via $\mu_s$, the **static** coefficient of friction. This result neglects rolling friction (and face impacts for the dodecagon), but demonstrates the inherent advantage to passing the rope over the top of the block. The definition for $\mu_d$ is the ratio of the work done in moving the block (and the attached rods) a given distance x, divided by the amount of work that would be required to lift the block a distance x/2 (recall that the crew moves twice as far as the block)

$$\mu_d = \frac{2T}{mg}. \qquad (2)$$

This definition gives $\mu_d$ in terms of the mass of the block, and the average tension, each of which is easily determined by the experiments.

The use of a rope to drag a large object is also well known to introductory physics students. The fact that one can apply the tension at an angle to the horizontal is a complication, but one can show that the minimum tension required to maintain a constant speed is less than $\mu_d mg$ and is minimized when the tension is applied at an angle above the horizontal $\theta$, such that $\tan(\theta) = \mu_d$. For the values expected for dragging ($\mu_d = 0.3$), one finds that the optimal angle is just a bit less than 17 degrees and that T = $\mu_d$ mg/(1.044), so that pulling at an upward angle is a minor effect.

To form the 4-sided block into the desired 12-sided polygon, the required rod diameter d, is the difference between D, the distance from the center of the block to a vertex, and b, the distance from the center of the block to the center of the appropriate face of the dodecagon of the same center to vertex distance D. By an extremely useful mathematical coincidence, three rods of that diameter give a surface of width 2d, which is exactly the length of a side on a dodecagon with the same center to vertex distance as the square. Using the relationship between the vertex distance D, and the distance to a face on a dodecagon, it is straightforward to show that

$$d = D\cos(\pi/12) - 0.707D = 0.366b. \tag{3}$$

Each set of three rods becomes a new "face" of the dodecagon, while the 8 remaining "faces" are formed by the line connecting the square vertex to the edge of the attached rod faces. Each face is of length 2d = 0.73b (see Fig. 1).

### III. EXPERIMENTAL METHODS AND RESULTS

Small scale experiments were conducted using a concrete block (40 cm x 19.6 cm x 20 cm, mass = 29.6 kg, density = 2040 kg/m$^3$). A set of three identical wooden "dowel rods" were attached to each face of the block parallel to the axis of symmetry using rope, transforming the square prism into a 12-sided polygon (a dodecagon). The desired rod diameter using Eq. (3) is found to be d = 3.66 cm. The diameter of the rods chosen were d = 3.80 cm, the closest standard size "dowel rod" available from the local hardware store (1.25, 1.5, and 2.0 inch). The details of the data collection can be found in prior work.[17] The most realistic tests were conducted on the infield section of a regularly maintained practice softball field on the campus of Indiana State University (ISU). Based on the total work needed to move the block a given distance in steady rolling motion across the infield, $\mu_d$ = 0.30 ± 0.05, is roughly equal to the best values stated for the small scale sliding experiments of Fall et. al.[6]

The large scale experiments were conducted in two configurations: (1) dragging on a sled, and (2) rolling as a dodecagon. The block used was a 1,000 kg sandstone block of dimensions 0.60 m x 0.60 m x 1.2 m (density = 2300 kg/m$^3$). Tension in the 26 mm hessian "pull rope" was obtained using a Horizon Mini Crate Scale (model SF-918 300kg/600 lb). The sled was constructed of wood, had dimensions 1.5 m x 0.7 m (surface area 1.05 m$^2$), and had a mass of 74.5 kg. The sled and block are shown in Fig. 2. The 12 posts used for the dodecagon were 100 mm in diameter (ideal would be 110 mm), 1.8 m in length, and had a total mass of 72 kg. The block is shown with rods attached in Fig. 3.

For the dragging experiments, the surface was prepared using 2 tons of river bed sand. The lubrication was provided by the rain that had fallen earlier in the day. The sled was first pulled across the sand, which required a tension of 340 N, giving roughly $\mu_d$ = 0.5, at least competitive with the values suggested for wet sand. The values obtained for the sled with the block on it were even better with an average value of 2470 N, giving $\mu_d$ = 0.25. It is not clear why the increase in the mass of the load was accompanied by such a large (factor of 2) reduction in $\mu_d$. The experiments were conducted on three separate occasions.

For the rolling experiment, the test area was covered in compacted, crushed sandstone, 50 mm maximum diameter. The sets of three rods were attached to each other using 200 mm Timberlock Bolts, and then attached to the block using 3.2 mm straining wire (22 m of it were needed), held under tension with 8 Medium Gripples (400 kg max load). The block with the rods attached is shown on the test surface in Fig. 3. Initial tests of the rolling were conducted without the pulling rope or tension gauges, and it was found that two men could move the block "relatively easily," and that person (S. Ward) could move it alone, with quite a bit of effort (see video link in ref. 18) over the 7 m distance. In the final configuration, the tension meter was added, and the block moved efficiently. As was observed with the small block, the tension required to initiate the motion TV, was larger than that required to maintain the motion, by about a factor of two. In particular TV = 490 N (about 0.05 mg), while TR = 270 N (about 0.03 mg), so that $\mu_d = 0.06$.

## IV. CONCLUSIONS

A novel method for moving large stone blocks has been presented. The method involves the use of wooden rods applied externally to the stone block to facilitate rolling the block, rather than sliding it. By attaching 12 identical wooden rods to the faces of the block, one effectively transforms the block into a dodecagon prism with very little added mass, much lower ground pressure, and with good cross country mobility. The results are qualitatively consistent with other researchers who attached wooden quarter circle "cradle runners" to full sized blocks, effectively turning them into cylinders. For the block of mass 1,000 kg, it was found that rolling gave $\mu_d = 0.06$, while sliding on wet sand gave a value almost five times as large at $\mu_d = 0.25$. Due to the 2:1 mechanical advantage inherent in the use of the rope being wrapped around the block, this means that the 9,800 N block can be rolled at a constant speed with an applied tension of less than 0.03mg = 290 N. A two-person team working at near maximum long-term sustained rate (150 W) could move at a speed of 0.5 m/s, so that the block would move at 0.25 m/s, much faster than even some of the fastest estimates used in calculating the time required to build the pyramids. The team of two could move the block the 2 km from the quarry to the Pyramid of Khufu in less than 2.5 hours. It would seem that some variation of rolling the blocks should now be considered to be among the "best" and most likely method used to move the stones for the great pyramids, or at least as a viable method for the next pyramid to be constructed by Earth Pyramid. The results certainly indicate that building the Great Pyramids was well within the human resource limits of ancient Egypt.

## V. ACKNOWLEDGEMENTS

The authors would like to thank the following for assistance: Erica O. West, Wilson Middle School, Terre Haute, IN for initial discussions and experiments with rolling blocks, Joseph Pettit, ISU Department of Geology for advice on the lashing of rods, Scott Tillman, University Architect for materials, Lori Vancsa, Office of Environmental Safety for safety equipment, and Stephanie Krull, Facilities Management Department for access to the experimental locations. In addition, the authors would like to thank the following for financial support, ISU College of Arts and Sciences (GG), ISU Department Chemistry and Physics (KW), ISU Summer Undergraduate Research Experience (SURE 2012). In addition the authors wish to thank Richard Dufton (master mason), and Peter Saunders, the owner of the quarry used for the large scale experiments at W. E. Leach Ltd, in West Yorkshire England.

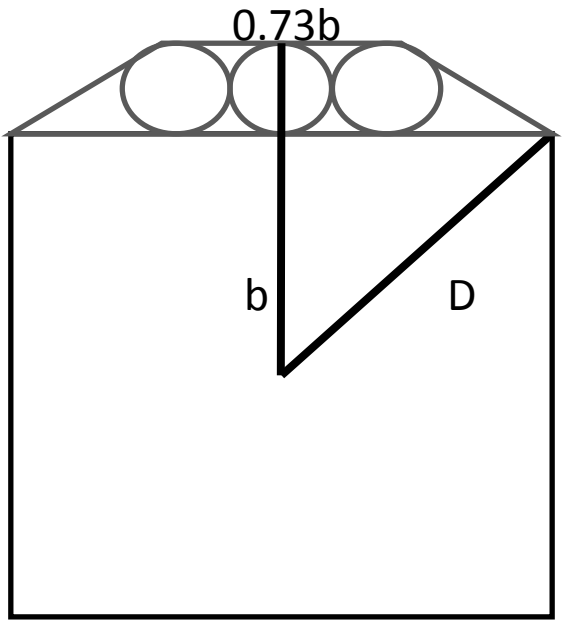

Figure 1. By adding three identical rods to one face of the square prism, that face is transformed into 3 identical faces of the dodecagon prism. The ideal rod diameter d is d= Dcos($\pi/12$) – b = 0.366 b, where b is from the center of the square to the face of the square, or half the length of one side of the square. Each face of the new dodecagon is 0.73 b in length.

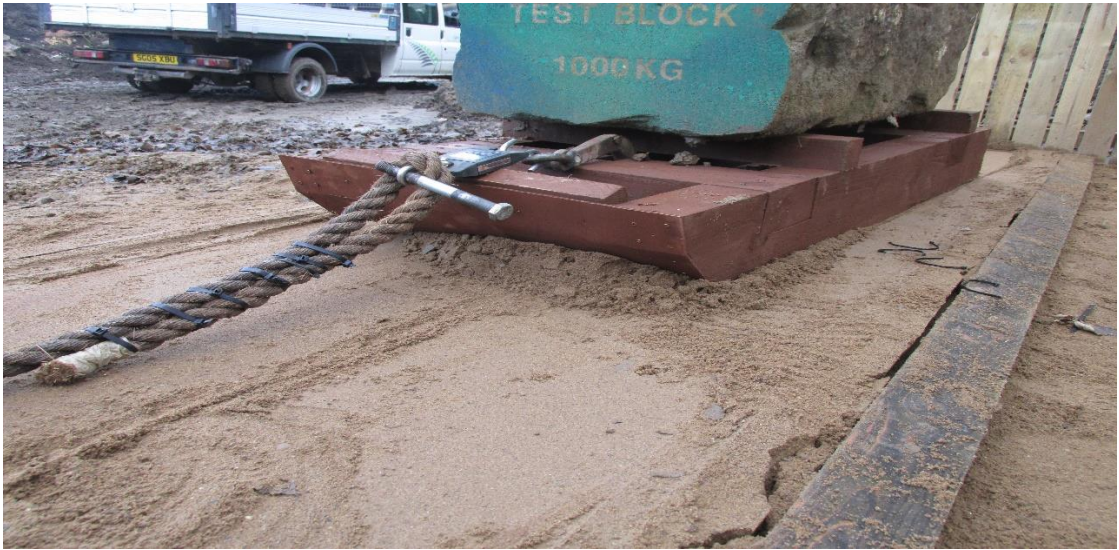

Fig. 2. Large block dragging experiment, showing the sled and the prepared area  Note the ridge of sand that builds up at the start of the motion, but remains at the same level once motion begins.

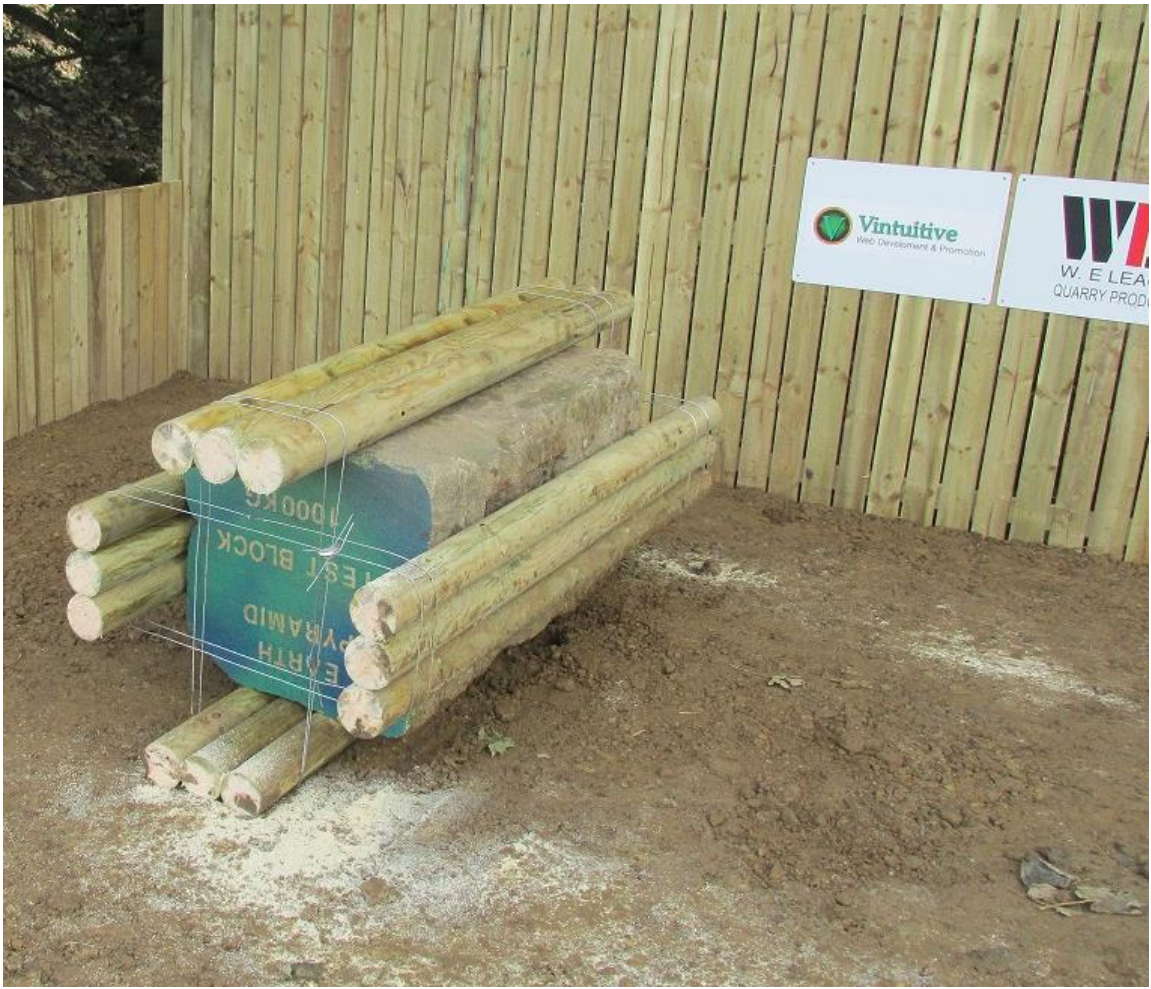

Fig. 3. Large block with posts attached. Note the degree of roughness of the test surface in comparison to the sand surface in Fig. 2.